# Network and Sentiment Analysis of Enron Emails

N. Belay, Graduate, Eastern Connecticut State University

N. Belay is with Eastern Connecticut State University, Willimantic, CT 06226 USA (e-mail: natnaelabe13@ gmail.com)

*Abstract*— The objective of the research was to analyze e-mails exchanged at Enron, a power company that declared bankruptcy in 2001 following an investigation into unethical operations regarding their financials. Like other researchers, we identify the "most important" employees and detect "communities" using network science methods. We find that the "importance" of a person depends on the centrality measure used; while the communities we detected resembled the formal organizational structure of the company. In addition, because previous work required that 10 e-mails be sent and received for an e-mail relationship to exist, we analyzed the effect of different "thresholds" on the results and found that results were very dependent on the "threshold" used. We also performed sentiment analyses on the e-mails to evaluate whether sentiment changed over time and found that the sentiments of the e-mails do not give insight into the financial wellbeing of Enron. Our results provide insight into how information flowed through Enron, who the key employees were, and e-mail sentiment before and after the crisis.

## I. Introduction

For an organization, understanding how employees communicate is very important. For example, it can help managers understand the culture of the organization better. Of the possible communication mechanisms, e-mail is the most popular. Managers may need to know how a particular message they sent is being passed around among the employees. Apart from managers in an organization, other people might be interested in e-mail communication from a social science perspective. In addition, analyzing e-mail communication can allow us to see the general sentiment of the e-mails being exchanged. Enron was chosen to perform this analysis because of the availability of corporate e-mails to the public. Enron was a power company that declared bankruptcy in 2001 following an investigation into unethical operations regarding their financials. By analyzing the contents of the Enron e-mails, and how e-mails are passed around, we can get an understanding of the social structure of a company and can identify the influencers of that employee network. The research also looks at whether the sentiment of the e mails corresponds to Enron's financial well-being.

The objective of the research was to analyze e-mails exchanged at Enron using concepts from Network Science and using Python programming to understand the informal organizational structure and the culture at Enron over time. To achieve our objective, we analyzed the Enron e mail corpus obtained from William Cohen of Carnegie Mellon University. We wrote scripts that traversed the e-mails, parsed out the necessary parts of the e-mails and conducted sentiment and network analysis. For the network analysis, we used different centrality measures to deduce the importance of a node (employee), and for the sentiment, we used TextBlob and Matplotlib to calculate and visualize the sentiment trend. Gephi was used as a visualization and centrality measuring tool. As part of this research, we provide three different Python modules, sentimentAnalysis.py, edgeListGenerator.py, and parseE-mails.py, on Github that can be used to conduct a similar analysis.

Our main findings are as follows:

· We find the ranking of the "most important" employees differ between the different centrality measures. We also find that the definition of a legitimate e-mail communication has an impact on the rankings of the employees.

· We find that the sentiment trend of the Enron e-mails does not reflect their financial troubles.

Researchers have investigated how the communication network can be analyzed to give insight into who the "most important" employees were, and how the structure of the network changes overtime. They have found that the communication network gets denser during the financial crisis of Enron [4]. In addition, they have found that different centrality measures give different results as to who the "most important" employee was [4].

The rest of this section is divided into the following sections: Section 2 contain background information and summary of previous related work, Section 3 describes the methods, Section 4 contains the results obtained from the methods, and Section 5 concludes the paper.

## II. Background and Related Work

### A. The Enron Company

Enron was a power company based in Houston, Texas and was co-founded in 1985 by Jeffrey Skilling and Kenneth Lay. Originally, the company started as a merger between two pipeline companies, Houston Gas and Omaha-Base InterNorth [1]. In 1988, deregulations for electrical markets were put in place and Enron became creative in order to increase profits in the newly deregulated market. The first action Enron took was to change their business model from "Energy delivery" to "Energy broker". That change allowed them to be market creators by bringing power demanders and sellers together. When a transaction takes place, Enron got paid the difference of the amount buyers paid and sellers requested [2]. The change in revenue model made Enron very successful financially and fed the culture of challenging limits.

In the 1990's Enron was one of the first companies to use the internet for their business operations [3]. The infusion of a strong energy company with a huge financial power and a presence on the internet built the confidence of investors which led their stock price to reach $90 per share and made the company worth over $70 billion [3]. The leadership of Enron actively promoted a culture of competition among employees and challenged everyone to match the standards that were being continually raised. Even though it seems that always striving to do better equates to success, there are some limitations that need to be respected in order to be within the limits of the law. Employees that worked at Enron have disclosed that the main focus of the Enron managers was to add value to their outputs, and to deliberately break rules to see how far they could get away with it [2]. As expected, this culture brought very serious issues in 1999. It was revealed that Enron had numerous partnerships with companies (known as special purpose entities) that they created [4], and it became clear that Enron was using those entities to separate the company's losses of $618 million and its debts in order to show

profits on their balance sheets. That action inflated Enron's profits which misled investors. The Security and Exchange Commission (SEC) began an investigation on October 31, 2001. The investigation revealed that Enron had a loss of $586 million for the previous five years. That news impacted the stock price which fell below $1 per share. Due to this financial loss, Enron filed for bankruptcy in December 2001 [4].

*B. Enron e-mail dataset*

As part of the investigation, the Federal Energy Regulatory Commission (FERC) released a database of Enron e-mails in May 2002 [4]. It is the only corporate e-mail corpus available to date for the public. The original database included 619,449 e-mails from 158 Enron employees which accounted for 92% of the Enron staff. The database contained the e-mail addresses of the sender and the recipient, date, time and body of the e-mail. This version of the database was not optimized for research and researchers have since restructured the data. The first modification was provided by Jitesh Shetty and Jafar Adibi from the Inter-service Intelligence (ISI). This version contained 252,759 e-mails from 151 employees. The e-mails in this version were placed in a MySQL table with different columns for different parts of the e-mail such as body and recipient. The version of data that was used in our research was from William Cohen from Carnegie Mellon University (CMU) [4]. This version contains 517,431 distinct e-mails from 151 employees [4]. The database size is 423 MB, and the e-mails are stored as text files. This version was chosen because the data integrity issues are solved.

*C. Mathematical and network science background*

A network is a relationship composed of nodes and links between those nodes. For example, a network could be employees and their relationships with each other. A node is an entity that is a subject or a target that performs an action in the network. For example, a node can be an employee that sends an e-mail in an organization. A link is a connection between nodes. For example, the e-mail sent from one employee to another links those subjects in the network. A network can be defined as a directed or undirected network. A directed network is a network where the links between the nodes indicate the directional flow of information. In a directed network, node can be a source of information and or a receiver of information (or both). An example of a directed network is the World Wide Web (WWW), where a web page has one or more links to other pages [5]. In our research, we used a directed network. Undirected networks are networks that have links which support the bi-directional transmission of information. An example of an undirected network is a power grid, where transmission lines transmits electricity in both directions [5]. In this case, a node would be a power plant and the link would be the transmission wires. An illustration of an undirected and directed network can be seen in **Fig 1** where the directed network has arrows at the end of the links and the undirected network has no arrows.

(i)

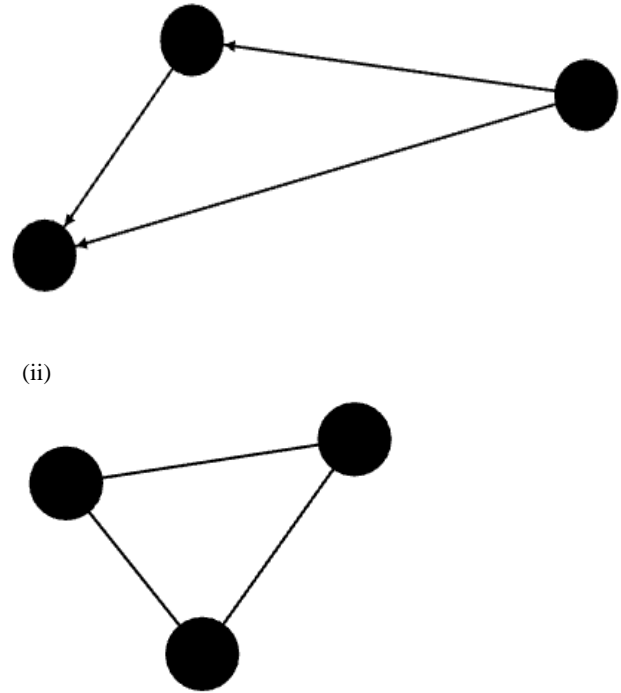

(ii)

**Fig 1. Example of a directed (i) and undirected (ii) network.**

*1) Overview of centrality measures*

Once the network is defined, numerous measurements can be made. Centrality measures are used to identify the most "important" or "influential" node(s) in a network, and include closeness centrality, degree centrality, betweenness centrality, eigenvector centrality, and PageRank. Because the centralities will measure different aspects of a node, the analyst has to decide which measures are appropriate for the question asked. The next section gives a background on the meanings and calculations of each of the above measures. Within a single network, there might be different nodes that are of highest importance depending on the centrality measure. Those nodes are illustrated in **Fig 2.**

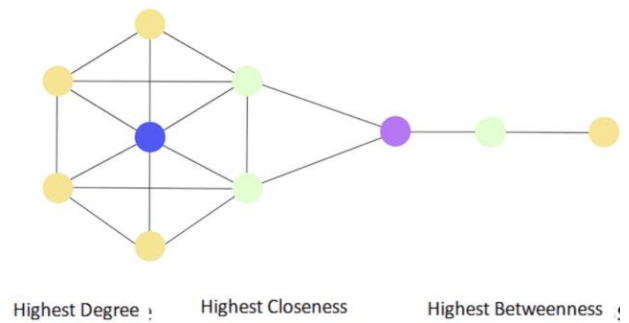

**Fig 2. An example of nodes with highest degree, closeness and betweenness centrality.** This Figure reproduced from https://docs.arangodb.com/3.2/Manual/Graphs/Pregel/[6]

*2) Degree centrality*

   *Description:* Degree centrality calculates how many nodes a single node is directly connected to. It is the number of connections a node has to other nodes. The node that has the highest degree centrality is the node that is connected to the most nodes [5]. In a directed network, there are two types of degree centrality measures. The first one is the in-degree centrality. It measures how many links point to the node. The other is out-degree centrality and it measures how many links point out to other nodes from that particular node. This measure is calculated for each node and a list of nodes in descending order of degree centrality is usually presented. The mathematical formula for degree centrality can be found in Ref # [7].

*Interpretation:* If one node has more direct connections than all other nodes, then this node may have more reach. From an organizational point of view, a person with the highest out-degree centrality disseminates information to a larger group of employees than anyone else. If an employee has a large in-degree centrality, they might be receiving a lot more information about the organization than anyone else. That information access can make them an integral part of an organization.

*3) Closeness Centrality*
*Description:* Closeness centrality is a measure of how close a node is to all other nodes [5], based on the shortest distance of a node to every other node. The shortest distance is the minimum number of links needed to connect two nodes. A node that has a small average distance from all other nodes will have a higher closeness. After calculating closeness centrality for one node, the process can be repeated for all nodes to get a value for each of them. The final values can then be sorted in descending order to rank the importance of each node based on its closeness centrality. The formula for calculating closeness centrality is given by Ref # [5] and is

$$C_i = \frac{n}{\sum_j d_{ij}}$$

where $C_i$ is the closeness centrality of node $i$ and $d_{ij}$ is the distance from node to node $j$, and $n$ is the number of nodes.
*Interpretation:* This measure can be indicative of an important person because it measures their reach to all other nodes. In an organizational setting, a person with a high closeness centrality can reach more of the employees in the organization with less relative effort. In other words, that person has the resources to get information from or to multiple employees quickly.

*4) Betweennes Centrality*
*Description:* Betweenness centrality is another measure in network analysis that calculates how many times a node is on the shortest path between different nodes [5]. Therefore, a node with high betweenness serves as an information broker for transferring data among nodes. This can be calculated as follows as in Ref # [5], using the formula

$$Z_i = \sum_{j,k \in V} \frac{\sigma_{jk}(i)}{\sigma_{jk}}$$

Where $Z_i$ is the betweenness centrality of node $i$, $\sigma_{jk}$ is the number of shortest paths from $j$ to $k$ and $\sigma_{jk}(i)$ is the number of those paths that go through node

*Interpretation:* Nodes with high betweenness have the power to control the exchange of information among employees because information passes through them to reach others quickly. These nodes can alter, block or facilitate information that is critical to the overall functioning of the network. If a network loses a node with a high betweenness centrality, then the information flow may become less efficient. In an organizational setting, an employee with high betweenness might be crucial for transferring important information such as company policies from one department to another.

*5) PageRank and eigenvector centrality*
*Description:* Eigenvector centrality and PageRank are very similar in how they calculate a node's importance based on the importance of the nodes that the node is connected to. However, they have their differences, and PageRank is considered to be a better measurement for a directed network than eigenvector centrality [8]. In eigenvector centrality, a node's centrality is calculated by how connected they are to nodes with higher importance. Therefore, they get their value from the properties of their connections. This centrality gives value to each node proportional to the sum of scores of its neighbors [8]. For eigenvector centrality, all the nodes inherently have 0 importance until they make a connection. PageRank is a similar concept but here all the nodes inherently have a certain β value, and the neighbor's importance is equally divided among its neighbors. Therefore, a node that is connected to more important nodes gets a higher rank. In addition, the formula of PageRank includes division by out-degree. This helps to keep nodes with high centrality from passing values to other nodes disproportionality by normalizing the value [8]. PageRank is a measure that Google uses to rank the relevance of webpages retrieved from a Google search. The formula for both eigenvector centrality and PageRank uses an adjacency matrix. An adjacency matrix is a matrix where the elements are defined as follows: if there is a direct link from node $j$ to node $i$, the value is 1; otherwise, the value is 0.

The eigenvector centrality $x$ of node $i$ is given by $x_i = \sum_{j \in V} A_{ij} x_j$ where $j$ is a neighbor of , $A_{ij}$ is an element of the adjacency matrix and $x_j$ is the eigenvector centrality of node $i$

The PageRank $y$ of node $i$ is $y_i = \alpha \sum_{j \in V} \frac{A_{ij}}{K_j} y_j + \beta$

where $\alpha$, $\beta$ are positive constants, $\alpha + \beta = 1$, $\beta > 0$, $k_j$ is the out-degree of a page (or node) $V$ is the set of all the nodes and $j$ is a neighbor of $i$ and is an element of $V_{(j)} \in V$ [5]. For more details, see Ref # [5]

*Interpretation:* When nodes are connected to other important nodes they will have the power to influence or get information from these important nodes and therefore are themselves important. In an organization, an employee that has more contact with managers may be considered important because of his/her link with important people. In this research, in addition to the other centrality measures, we also used PageRank, which to our knowledge has not been used for detecting important people at Enron. An example PageRank is shown in **Fig 3** where the sizes of the nodes correspond with their respective PageRank value.
Figure 3. An illustration of nodes with their sizes showing their PageRank rankings**.**

*6). Community detection*
*Description:* Community detection is another concept in Network science that aims to identify subsets of nodes that are more connected with each other than they are connected with other subsets. One method to identify communities **is** Modularity [5]. In this method, the network of nodes is divided into n random partitions [5]. Then, nodes are moved from one partition to another while recording the change in Modularity. If the Modularity of the network increases when a node is moved to a different partition, then it suggests that the node belongs in that partition. The community structure that results in the highest Modularity measure is considered to be the optimal partition

[5]. Refer to Ref # [8] for the formula to calculate Modularity.

*Interpretation:* Because community detection identifies subsets of nodes that are more connected with each other than they are connected with others, the process can be used to identify "real life" e-mail communities at Enron, which will help understand the company's organizational structure and culture.

*7) Sentiment analysis*

*Description*: Sentiment analysis is an area of Natural Language Processing (NLP) that quantifies the emotional content of text. Generally, sentiment is categorized as positive, neutral or negative. One of the tools that can be used for conducting sentiment analysis is a Python module named TextBlob which also provides a simple API for conducting text analysis tasks such as part-of speech tagging, noun phrase extraction, classification, and more. TextBlob uses a predefined set of words in a dictionary(lexicon) that are assigned subjectivity, intensity, polarity (sentiment values), position in a sentence (part-of-speech), and a short description (called sense). A single word might have multiple sense, therefore complicating the sentiment calculation. This lexicon is an open source XML document that can be found in the following link https://github.com/sloria/TextBlob/blob/eb08c120d364e908646731d60b4e4c6c1712ff63/textblo b/en/en-sentiment.xml [9].

*Calculation*

Single Words: At the core, Text Blob uses averaging to calculate the polarity of words. For instance, the word great has four senses (possible meaning), as a result having four different polarity values in the lexicon [9]. To get a concrete answer, the four polarity measures are averaged together to result in a polarity of 0.8 [9].

| Word  | Polarity | Subjectivity | Intensity | Sense |
|-------|----------|--------------|-----------|-------|
| Great | 1.0      | 1.0          | 1.0       | "very Good" |
| Great | 1.0      | 1.0          | 1.0       | "of major significance or importance" |
| Great | 0.4      | 0.2          | 1.0       | "relatively large in size or number or extent" |
| Great | 0.8      | 0.8          | 1.0       | "remarkable or out of the ordinary in degree or magnitude or effect" |

*Negations:* Whenever TextBlob encounters a negation, it multiplies the polarity by $-0.5$ leaving the subjectivity and other measures unchanged [9]. For instance, the polarity of "not great" would change to $-0.4$ because $0.8 \times -0.5 = -0.4$ [9].

*Modifiers:* Modifiers, like very, will have an impact on the polarity and subjectivity of the of the word. In case of modifiers, the intensity of the modifier is used to multiply the polarity and subjectivity of the of the word [9]. For instance, "Very Great" would have a different sentiment than just "great". Here the intensity of "very" is 1.3 [9]. Thus, both the sentiment and the subjectivity would be multiplied by 1.3 [9]. Since both the subjectivity and polarity have a value that ranges between $-1$ and $+1$, any value that is outside the boundary would be set to either numbers depending on the sign of its value. Therefore, "very great" would have a polarity of 1 (even though its true value is 1.04) and a subjectivity of 0.975 [9].

*Negations and modiiers:* it is possible that there are phrases that have both modifiers and negations. For instance, the word "not very great". In this case, the calculation is somewhat similar to the above calculations. The polarity would be multiplied by -0.5 for the negation [9]. But then, for the modifier, instead of multiplying the intensity as it is, the inverse intensity is multiplied for both the polarity and subjectivity [9]. Therefore, the polarity of "not very great" would be the result of $\frac{1}{1.3} \times -0.5 \times 0.8 = -0.31$. The subjectivity would be $\frac{1}{1.3} \times 0.75 = 0.58$ [9]. Single words and unknown words: When TextBlob encounters a single letter word or a word that is not in the lexicon, it will simply ignore it by assigning a zero value. For longer sentences, TextBlob averages the values of the sentiments for each words or phrases.

*C) Previous analysis of Enron e-mails and related work*

Klimt and colleagues worked on automated classification of e-mails to user created folders based on Natural Language Processing (NLP) [10], using the CMU version of Enron's e-mail database. The research objective was primarily to classify, by folders, e-mails by looking at the "To", "BCC", "CC", and "From" fields, using Support Vector Machines (SVMs). They conducted their analysis by feeding e-mails to the SVM with each of the 4 fields listed above separately and all the fields together as a bag of words [10]. They then calculated the F1 score (a measurement based on the precision and recall) and determined which fields are most important for automated e-mail classification. Their research found that the performance of the classifier depends on the folder creation strategy of the user. The more folders a user has the more accurate the SVM classification. The limitation of this research is that it is specific to the Enron e-mail dataset and the performance of the SVM depends on the organization of a user's folders, which are determined by each user.

Diesner and colleague studied the Enron e-mail data set to explore the communication network [4]. Their primary goal was to compare different network science measures before and after the crisis at Enron occurred. They specifically compared October of 2000 to October of 2001, which was around when the Enron crisis broke out. They used the dataset provided by Shetty and Abidi from ISI but added a layer to the dataset by introducing information on the positions of the Enron employees. They named this instance of the data set the Enron CASOS database. They segmented the data into monthly e-mail exchanges so that there is a detailed view of the change of communication that occurred within the years. They fed the data to an Organizational Risk Analyzer (ORA), so that they can visualize the change in the clustering of e-mail exchange among agents.

They defined a relationship between agents as an e-mail being sent between them at least once. As expected, October 2000 had a less dense network as compared to October 2001. They calculated different network analysis measures for the two time periods to identify the top five agents. They looked at the measures closeness centrality, betweenness centrality, eigenvector centrality, in-degree centrality, and out-degree centrality. Each of those measures came up with different employees as having high measures and the employees with the highest measures also changed their standing between 2000 and 2001. For instance, they found that in October 2000 the person with the highest closeness centrality was the manager W. Stuart. However, in 2001 an employee named S. beck had the highest closeness centrality. An important limitation with [4] is that they used a single e-mail as a criterion for an e-mail relationship. Therefore, e-mail relationships may be based on error messages, spams or one-time e-mail exchanges that are not representative of the true e-mail relationship structure. Our research compared and analyzed different

e-mail thresholds to define an e-mail relationship. In addition, we used at PageRank algorithm, which is considered better than eigenvector centrality for directed graph.

Shetty and colleagues used a different methodology to deduce the most important agent [11]. They used the order (peace) of a network in order to see what node disturbs the overall graph the most when removed. For this purpose, they used Graph Entropy to calculate the order of the Graph. They used the e-mail corpus provided by ISI which is stored in a MySQL database. They used one e-mail between agents in order to establish an e-mail relationship. After that, they carried out two distinct iterative processes of removing each node from the graph and recording the change in Graph Entropy. One of the processes was concerned with removing nodes and links that are directly connected to the node, links that are at a distance of one link from each other. The second process was removing each node and its directly linked links and in addition links that have used the node as a transit, links that are at a distance of two links from each other. The node that had the highest change in entropy was the most important. The top five nodes with highest disturbance were recorded for both processes. In length=1, they found out that the most important person was Louise Kitchen, Enron ex-president. On length=2, they found out that Greg Whalley, Enron ex-president, to be the most important. They concluded that length=2 gives a more realistic answer because the reach of an employee usually stretches more than their immediate contacts. The limitation of this research is that they did not consider removing links that are more than a length of two. The reach of some employees might be more than two links and [11] ignores those employees. They also did not provide any justification to why they chose a length of two.

Joshua and colleagues studied the methods of automatic community detection using e-mails gathered from HP labs [12]. They used the Girvan Newman algorithm to identify communities. Their research was based on 1 million e-mails gathered within a period of 2 months from November 25, 2002 to February 18, 2003. However, that dataset was cleaned by removing messages sent to a list of more than 10 recipients, as these e-mails were often lab wide announcements, which consisted of 185,773 e-mails. From the e-mail content, only the "To" and "From" fields were used in order to facilitate the identification of communities while keeping the privacy of the agents. The definition that they applied to define an e-mail relationship was that a person had to send at least 30 e-mails and get at least 5 back from the same person. That definition helps to create a relationship network where the communications are legitimate. They used the non-local process of partitioning a network into subsets by using betweenness centrality. This process helps visualize any nodes that have links to other subsets and suppresses them in that subset if they have more links in the subset that outside of it. The method involves repeatedly identifying intercommunity links and removing them until the giant component is resolved into many separate communities. Their algorithm stops removing links once the number of vertices in a subset equals 6. That is because to have a viable component of more than one community, there needs to be at least 6 vertices in both communities combined connected to each other by a single link. If the link is removed, the components vanish, and the vertices become part of the same community. In partitioning a large connected component of vertices, instead of using every vertex as the "center" once, they cycled through the network randomly by choosing M centers instead of all vertices becoming centers. This process was carried out until the betweenness of a link exceeded the betweenness of the leaf node. They then remove the link with high betweenness (meaning it connected components). This process is repeated until the entire graph was separated into communities. This process keeps a running total of how many times each vertex appeared in a community on each round. If there were $n$ rounds and a vertex appeared in a community $n$ times, it suggests that the particular vertex belongs in that community. If the running total of the vertices is smaller than $n$ it means that it is also related to other communities. In this research, they only used the "To" and "From" fields from the e-mails to construct the network structure. In our research, in addition to the "To" and "From" fields, the "CC" and "BCC" fields were also be used to create a more accurate representation of Enron. The community detection algorithm that was used was the Modularity.

Page and colleague ranked the importance of webpages using the PageRank algorithm [13]. They built a repository of 24 million webpages through complete crawling and indexing as part of the Stanford WebBase project. They converted the URLs to integer ID's, stored the links in a database with the integer ID serving as a unique identifier, and removed dangling (dead-end) links. As part of the PageRank calculation, every node, in this case every web page, gets a rank. PageRank runs through the database iteratively until it converges and assigns the natural rank of a webpage. PageRank is also related to the "Random Walk" probability, which is the probability that a random walker, or a random user, ends up at a certain webpage. They used PageRank for page searching, similar to what Google does. Following a title search, they are able to get an accurate list of pages that are logically related to the search term. This paper demonstrates PageRank, which was used in our research.

III. METHODS

*A) The e-mail corpus and reproduction of previous research*
The e-mail corpus is downloaded as a set of folders and text files using a Windows Personal Computer. The network diagram is based on the e-mail corpus where a node is an agent and a link is an e-mail relationship between agents.

Our first step was to reproduce the results obtained by [4]. In that effort, we first retrieved all the edge lists that were included in the dataset. We used the Python programming language to read through the e-mail messages and retrieve the different sections of each e-mail. To be added to the edge list, we required the individuals in the "From", "To", "CC", and "BCC" fields to have a folder in our dataset. The filtering process was conducted using an SQL analysis in Microsoft Access. Once that filter was set, the number of employees we had data on was 146.

We then inputted the filtered list into Gephi. We also added the positions of the employees that were ranked. For the position data, we used an external source, from University of Edinburgh School of Informatics, which had the employee names along with their positions [14]. We parsed our dataset and performed another SQL analysis in Microsoft Access that counted the number of employees that correspond to each position. Finally, we produced communication networks using Gephi by inputting the edge list. PageRank algorithm was used as a criterion to size the nodes. For the analysis, only the "sent", "_sent_mail", "sent_items" e mail folders were utilized

*B) Sensitivity to alternative definitions of e-mail relationships*
We changed the number of e-mails that were required to be exchanged between two employees (threshold). We performed similar analysis as above (reproduction of previous research) by changing the definition of a legitimate e-mail communication between two employees (threshold). We defined the thresholds at different points, 0, 5, and 10 e-mails sent from one employee to another. Gephi's filtering function was used to filter for different edge weight's by keeping the minimum degree value at a constant value of one. The edge weight filter was a subquery of the degree number filter in Gephi. The different rankings of the employees from October 2000 where plotted.

*C) Sentiment analysis*

To conduct the sentiment analysis of the e-mails, a Python package called TextBlob was used. Once the files are located and the body of the e-mail extracted, the TextBlob package was invoked to give a numerical representation of the sentiment of each e-mail. For this research, in order to know the change of sentiment over time, the e-mail sent date was parsed out using Python programming as the sentiment of the same e-mail was being calculated. An average of the sentiments for each month was calculated and plotted in a graph using the Matplot package of Python to visualize the change in sentiment over time.

Two different graphs were produced as part of the analysis. The first line graph showed the time series trend of the sentiment on the organizational level. From it, we statistically compared the average sentiment values for October 2000 and October 2001. The second line graph showed a time series trend of the sentiment for every user.

*D) Community detection*

For our analysis, we used the modularity community detection algorithm that is provided by Gephi. It calculates the modularity values of each nodes and assigns them to a class. Thus, a class corresponds to community. We then color coded the different classes to create a visual representation of the different communities. Within Gephi, ForceAtlas2 with a Strong Gravity
was used as a layout option for the network.

## IV. RESULTS

*A) Development of Python Module for Enron e-mail analysis*

As part of this research, we have developed three Python modules to help researchers analyze the Enron e-mail dataset. They are available in the following Github repository: https://github.com/NatnaelA/thesis. They are described as follows:

*parseEmail.py:* This module is the foundation of the all the other modules because it helps to extract different values from the e-mails. The sections extracted are "From", "To", "CC", "BCC", "Subject", "Body", and "Date". The module allows the user to specify which of those fields they want to recover. Multiple selections are also supported. From this module, a user can select to parse the email of a single e-mail, a single employee folder, or all the employees included in the data set. A user can also specify the start and end date of the e-mails to focus on.

*generateEdgeList.py:* This module builds on the parseEmail.py module to use the "From", "To", "CC", and "BCC" fields to create an edge list. It has the capability to generate an edge list for a single email, for a single employee, or all the employees in the data set. A user can specify the start and end date of the e-mails to focus on.

*sentimentAnalysis.py*: This module builds on the parseEmail.py module; however, it has additional components. Like the other modules, it allows the calculation of the sentiment for a single email, a single employee or all the employees included in the data set. A user can also specify the start and end date of the e-mails to focus on. In addition, this module completes execution by returning a sentiment trend graph. The user can choose to select a line graph or a box plot to see the time series trend of the sentiment at Enron. The default is a line graph.

*B) Reproduction of previous research*

*1) Enron e-mail data corpus*

The specific dataset used has a significant impact on the results that will be revealed. Analyzing the dataset gave us a strong foundation to interpret our results. In an effort to reproducing earlier research, we summarized the number of filtered employees associated with each position and compared it to other data sets. The reason for filtering is that we wanted to get an accurate representation of the conditions inside Enron. The unfiltered dataset would have e-mails sent to outsiders, like family and friends, which would add noise to our analysis. We found that there were similar positions across the datasets. However, there are differences in the count, and some datasets had unique positions. That is mainly due to the existence or nonexistence of certain employees from the data sets. Table 1 shows the summary statistics of the employee.

Table 1: Count of Enron employees at each position from different dataset

| Position | Our data | Ref #4 |
|---|---|---|
| Analyst | 1 | 10 |
| Assistant | 2 | 0 |
| Associate | 0 | 5 |
| Attorney | 2 | 0 |
| CEO | 4 | 4 |
| Director | 15 | 27 |
| Employee | 34 | 69 |
| Head | 0 | 2 |
| In House Lawyer | 0 | 3 |
| Lawyer | 2 | 0 |
| Legal | 4 | 0 |
| Manager | 18 | 31 |
| Managing Director | 2 | 6 |
| President | 4 | 4 |
| Senior Manger | 2 | 0 |
| Specialist | 1 | 9 |
| Sr. Specialist | 0 | 17 |
| Trader | 23 | 9 |
| Treasury Support | 0 | 2 |
| Vice President | 25 | 29 |

In addition, we were motivated by previous research by Diesner and Carley to see how the number of e-mails sent progresses through time [4]. We conducted two analysis where one included all the sent e-mails and the other filtered for only the Enron employees that we have data for. As expected, the range of the number of e-mails is larger for all the sent e-mails, with a peak of 8,000 e-mails, as compared to the filtered dataset, with a peak of 1,000. Diesner and Carley had a peak for all the e-mails at 30,000 and a peak of about 5,000 for a subset of employees [4]. **Fig 1A** (see appendix) shows the distribution of the e-mail count for all employees and Fig 2A (see appendix) shows a distribution for the filtered set. As expected, Fig 1A has two major peaks on April 2001 (the month were analysts started to question about Enron's earnings and transparency) and a second one on October 2001 (the month where the investigation into Enron began). Unlike our results, [4] had a result were the October peak was significantly higher than the April one. This difference in results can be attributed to the difference of dataset that was used. On the filtered data set (Fig 2A), the difference between the April peak and the October peak becomes more evident with the April one being significantly higher.

*2) Centrality measures*

As an effort to check the validity of the results presented by [4], we performed a similar analysis where we calculated the centrality measures of the employees and we ranked them. We used Gephi to construct the network and measure the centralities of the employees and rank them in descending order of their values. Our results, as shown in Table 1A (see appendix), Table 2A (see appendix), and Table 3A (see appendix), show that using different centrality measures leads to different results regarding who is the most important person. For instance, for October 2001, Eigenvector Centrality has John Lavorato, the CEO, as the employee with the highest value, but Betweenness Centrality had Louise Kitchen, the President, as the most important employee. This is understandable because Betweenness Centrality measures how often an employee, is in the middle of the shortest path between other pairs of employees. On the other hand, Eigenvector Centrality gives a value to a node according to the importance of the other nodes that it is connected to. For this step, we defined an e-mail communication (threshold) as at least one e mail sent between employees. Refer to sections 3.2 and 4.3 for more information on thresholds. In addition, our results differed from [4]. Even though the two analyses had common individuals, there are a few that are unique to either dataset. Since most centrality measures give value to a node relative to the other nodes, the non-existence of a node will affect the centrality measure of the other nodes. For instance, eigenvector gives a value to a node depending on the value of the node that it is connected to. Thus, if A has a higher value (importance), B will also have a higher value if it connects with A. However, if A is not in the dataset, the value of B will be reduced which will then affect its ranking.

*3) Network Structure*

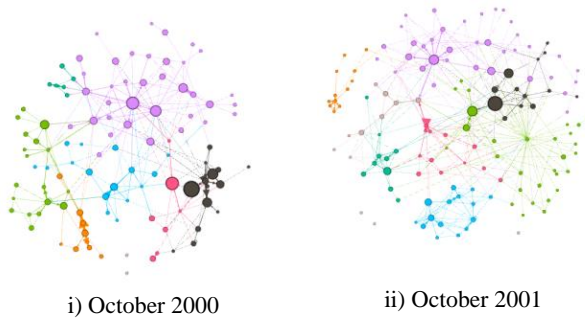

i) October 2000      ii) October 2001

**Fig 4: Enron communication network in October 2000 (i) and October 2001 (ii)**

In fig 4, The sizes of the nodes differ due to the ranking of the nodes according to the PageRank algorithm. Consistent with [4], it can be seen that the network in October of 2001 is denser and more connected than the network for October 2000. In October 2000, the number components were 84. In October 2001, the number components were 72. A component is a set of nodes where all pairs of nodes have paths connecting them. Thus, as the number of components increases, the graph becomes more disconnected because there will be numerous subgraphs. For a cohesive network, the total number of components should be lower. Thus, in October 2001, it can be seen that the number of components was reduced indicating that the more nodes are becoming part of a single component, as a result more cohesive. This is understandable because employees would be communicating more during a crisis.

*C) Evaluating the choice of threshold on node importance*

Because previous work by Shetty and Adibi required that 10 e-mails be sent and received for an e-mail relationship to exist, and Ref #4 had none, we analyzed the effect of different "thresholds" on the results and found that results were very dependent on the "threshold" used. A threshold is the minimum number of e-mails that need to be sent to define a link. From Network Science perspective, this threshold is a lower bond for an edge weight of the network. We were interested in comparing the effect of different thresholds on the centrality measures and thus on the ranking of the employees. The result was that the definition of a threshold had an impact on the centrality measures of employees, and as a result altered the ranking of importance. The different thresholds used were 0 (no threshold),5, and10. Gehpi's filtering functionality was used to apply the threshold. The edge weight filter was a subquery of the degree number filter. This would mean that first, all the nodes that have at least a degree value of one will be filtered (send at least one e-mail). Then out of those results, the edge weight filter will choose the edges with a certain value of an edge weight we have specified. The motivation behind this is to produce a real relationship because if employees send a higher number of e-mails among each other, there is a higher probability that there is a professional e-mail relationship between them. This threshold filters out connections where employees only send a few e-mails to each other on certain occasions and where there is no real professionPal relationship between them. The 0 threshold was performed in the previous section (section 4.2) while reproducing [4]. Table 4A shows how the centrality ranks changed when the threshold was 0, 5, and 10. The maximum edge weight (most e-mails sent to an employee) for October 2000 was 44 by Eric Bass, Trader, and for October 2001 was 89 by Jeff Dasovich, Government Relations Executive.

*1) Threshold impact on ranking*

To visualize how the rankings of each employees change with threshold, we plotted a graph that shows the progression of the ranking across thresholds. Fig 3A illustrates that different threshold values result in different rankings about employees. By increasing the threshold, some employees were unranked for that specific threshold because they were either filtered out by the requirement or they were not ranked high enough for that threshold because most of their links disappeared. Once they are unranked, they do not reenter the top 5 for another threshold.

*2) Impact of threshold on the network*

Threshold change also had an impact on the network structure. The change resulted in the disappearance of nodes from the network, which impacted the centrality measures of the other nodes. As a result, fig 4A shows that as the threshold increased the network becomes less dense and there are multiple island networks that resulted from the disappearance of edges in the filtering processes. These characteristics was also seen in October 2001.

*3) Community detection*

We use Gephi for community detection based on Modularity. All nodes with the same colors are in the same community. From our analysis, we discovered that the community formed was similar to the formal organizational hierarchy. For instance, **fig 5A** shows the overall community structure with a closeup to a specific community. The source data was the filtered edge list of October 2000. The employees emphasized worked in the same field: the legal issues. **Table 2** shows the names and the positions of the employees represented in **fig 5A**.

TABLE 2: LEGAL DEPARTMENT EMPLOYEES AND THEIR POSITION

| Employee | Position |
| --- | --- |
| Mark.haedicke@enron.com | Managing Director-Legal Department |

| louise.kitchen@enron.com | Creator of Enron Online- Worked closely with Legal |
|---|---|
| marie.heard@enron.com | Lawyer |
| mark.taylor@enron.com | VP and General Counsel |
| stacy.dickson@enron.com | Counsel |
| stephanie.panus@enron.com | Specialist Legal |
| susan.bailey@enron.com | Specialist Legal |
| susan.pereira@enron.com | Manager in Trading |
| tana.jones@enron.com | Specialist Legal |
| carol.clair@enron.com | Gen Counsel Assistant |
| sara.shackleton@enron.com | Gen Counsel Assistant |
| Michelle.cash@enron.com | Assistant General Counsel |
| Gerald.nemec@enron.com | Attorney |
| Janette.elbertson@enron.com | Assistant to the Director -Legal Department |
| Kay.mann@enron.com | Head of Legal Department |
| Elizabeth.sager@enron.com | Vice President and Assistant Legal Counsel |
| dan.hyvl@enron.com | Employee- Legal department |

*4) Sentiment analysis*

After observing that the number of e-mails sent was much higher around the crisis, we were motivated to conduct a sentiment analysis to analyze if there is a similar change in the sentiment of the e-mails. By using TextBlob, we parsed and analyzed the body of the e-mails by keeping track of the sentiment scores. TextBlob compares each word in the e-mail with a pre-defined dictionary of words (lexicon), included with TextBlob, that have a sentiment scores associated with them. The overall sentiment of a longer sentence is average of the words or phrases that TextBlob recognizes. We use TextBlob for sentiment analysis, as described in Section 2.4.7. We used matplotlib to plot the sentiments. Two graphs were constructed: the first one being a sentiment graph on the organizational level, shown in **fig 6A**, and a second one was a graph on a employee level, shown in **fig 8A**. On the organizational level, the trend fluctuated around a single point for a significant amount of time which made our result inconclusive, however; on the employee level, there are users who had strong sentiment (either positive or negative). For instance, Bradely Mckay, Contractor, had the highest average positive sentiment in December of 2000. The increase in sentiment on December 2000 could be related to Jeff Skilling replacing Kenneth Lay as the CEO. In the same month, the stock hit a 52-week high of $84.87 [15].

We also performed a similar analysis on the sentiment of the e-mails that are filtered (shown in **fig 7A**). These e-mails were sent between employees that we had data for.

Comparing the October 2000 and October 2001 sentiment results in both **fig 6A** and fig 7A gave a counter-intuitive result. We first assumed that as the crisis approached, the sentiments of the e-mails would decrease. However, it can be seen that the average sentiment in October 2000 was lower than the sentiment in October 2001. Quantitatively, the sentiment for October 2001 was 72% higher than October 2000 with a P-Value less than 0.05 (the result is not due to chance). Fig 5 shows the difference in the average sentiment between the two years. Thus, it was concluded that the e-mail sentiments did not represent the financial crisis that Enron was experiencing.

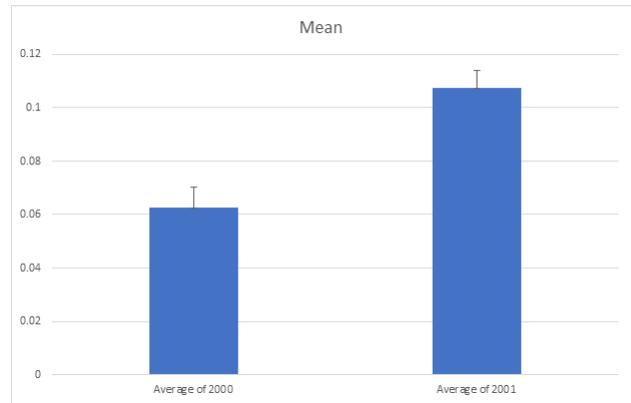

**Fig 5:** Average Sentiment for October 2000 and October 2001

Noticing this result, we compared the sentiment trend along with the trend of the stock price in **fig 6**. Our comparison produced that there was no correlation between the sentiment of the e-mails and the stock price.

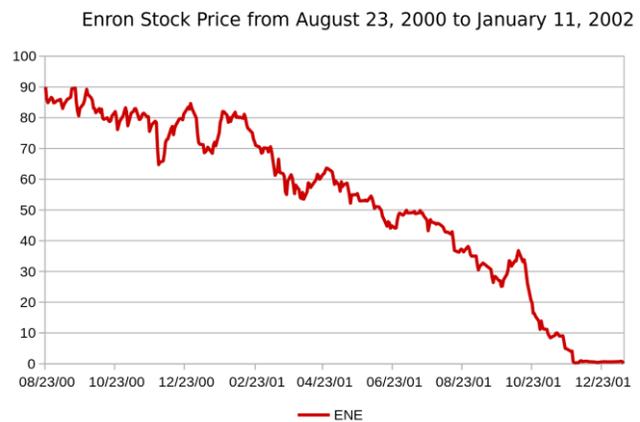

**Fig 6:** Stock trend for Enron. This Figure reproduced from https://www.begintoinvest.com/enron-stock-chart/ [16]

Relating the sentiment analysis result to the result of the number of e-mails sent, it can be seen that there is also no relation. Meaning, the increase the number of e-mails sent during a period does not give us an indication of the trend of the sentiment. For instance, in **fig 1A**, there is a peak of number of sent e-mails in April 2001. However, looking at the organizational sentiment graph, **fig 6A,** the change in the sentiment is not noticeable, it only changed by 0.025 going from 0.075 to 0.1.

## V. LIMITATIONS

The research has some data integrity issues regarding the e-mail corpus. Since the corpus was not built to specifically answer the questions raised in this paper, some e-mails might be irrelevant for the analysis. For instance, some e-emails might be spams that should not be representative of the organizational structure. Additionally, the e-mail corpus only contains emails of 92% of the Enron ex-employees. This limits the results of the research not to be 100% accurate. In addition, our analysis is done on Enron sent e-mails collected between 1998 and 2002. Due to that, we would be missing received e-mails and other received e-mails that have been moved to another folder. Finally, the TexBlob module uses averaging to calculate the polarity of the e-mails and thus lacks context. Therefore, some e-mails might have inaccurate sentiment readings.

## VI. Conclusion

E-mail is one of the most popular communication mediums for employees in the work place. Using those e-mails, one can construct a communication network structure that gives insight into to how information flows through the organization and the overall culture. With this motivation, we conducted a network and sentiment analysis where we tried to see who the "most important" person was, the change in sentiment of the e-mails sent at Enron overtime, and how changing the definition of an e-mail relationship affected the results. Like previous researches, we found that different network science measures gave different people as the "most important". We also found that the network science measurements were sensitive to alternative definitions of e-mail relationships (thresholds); we used thresholds of 0, 5, and 10. In addition to the threshold, the calendar year in consideration had impact on the ranking of the employees. During the crisis, October 2001, the network was denser indicating that employees were communicating more. Due to the change in the character of the network, the rankings of the employees also changed between October 2000, a year before the crisis, and October 2001. As a result, researchers should be careful when choosing threshold values and the calendar year because they would have a significant impact on their result. Community detection using Gephi has allowed us to distinguish six different communities: Senior Manager for Traders, Executive Team, Traders, ENS Gas East, ENS Gast West, Associates in Trading, and the Legal team. Finally, we find that a sentiment analysis of the e-mails does not give indication in to the financial status of Enron. The average sentiment in October 2001 was actually higher than October 2000 with a P-value of less than 0.05. The sentiment trend also does not correspond to the trend of Enron stock price.

## Autor Biography

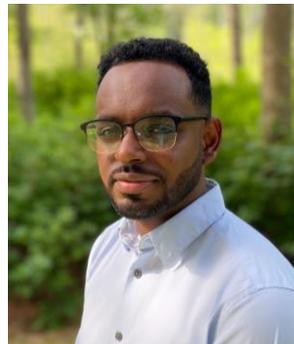

**N. Belay** (Graduate, Eastern Connecticut State University). I was born in Addis Ababa, Ethiopia on January 13th, 1997. I have a bachelor's degree from Eastern Connecticut State University, which is located in Willimantic, CT USA. I graduated in May 2020 with a dual major in computer science and business information systems. He currently has been working as a Technical Program manager at Google since May 2022. He is based out of Cambridge, MA office. At Google, he is responsible for leading the Android Jetpack program, and the Android API Council. Before Google, he worked as a Program Manager at PTC